# Intrinsic Calibration of Molecular Alignment Using Rotational Echoes

Dina Rosenberg[1,2] and Sharly Fleischer[1,2*]

[1]*Raymond and Beverly Sackler Faculty of Exact Sciences, School of Chemistry, Tel Aviv University 6997801, Israel.*
[2]*Tel-Aviv University center for Light-Matter-Interaction, Tel Aviv 6997801, Israel*
*\*Email: sharlyf@tauex.tau.ac.il*

Abstract: We propose and experimentally validate the use of rotational echo responses for obtaining the degree of molecular alignment induced in a gas. The method is independent of various parameters that are hardly accessible in most experimental configurations such as the effective length of interaction and the gas density as it relies on the intrinsic, self-contained dynamics of the rotational echo response.

---

### *Introduction:*

Coherent control of molecular rotations has been thoroughly studied and vastly utilized in the last three decades. Motivated by obtaining spectroscopic signatures from the molecular frame, researchers have invented a large and sophisticated toolbox for controlling the angular distribution of gas phase molecules via their interaction with strong laser pulses [1–5]. In fact, the field of rotational control has evolved in parallel to and from rotational coherence spectroscopy that aims to obtain accurate measurements of molecular rotational coefficients, from which the molecular structure can be deduced [6,7].

The basic physics of laser-induced rotational dynamics of linear molecules is well understood: an ultrashort laser pulse interacts with the molecules via their anisotropic polarizability and applies an effective torque that rotates them toward the polarization axis of the pulse. Shortly after the interaction (tens – hundreds of femtoseconds later) the rotating molecules become aligned, i.e. with their molecular axes preferentially lying along the polarization axis of the excitation pulse (taken as the z-axis). As the molecules continue to rotate freely, they dephase and regain the isotropic angular distribution shortly after [8,9]. However, the quantum mechanical nature of molecular rotation imposes quantization of the angular momentum and of energy levels with $E_{J,m} = hBcJ(J+1)$ where $J$ is the rotational quantum number, $B$ is the molecular rotational coefficient in $[cm^{-1}]$, $c$ is the speed of light and $h$ is Planck's constant. This quantization manifests in periodic recurrences of aligned and anti-aligned angular distributions of the ensemble throughout the coherent evolution of the ensemble with a period given by $T_{rev} = 1/2Bc$, also termed "the rotational revival period".

*Quantifying the degree of alignment:*

In many experiments it is highly desirable to quantify and report the exact degree of angular anisotropy induced in the gas, and in some cases even crucial. This may be achieved by Coulomb explosion imaging [10–13] that provides access to the actual angular distribution of molecular axes however, applies to low density gas samples with single or up to few tens of molecules at the interaction region. However, for experiments and applications such as HHG [14,15], collisional cross section measurements [16–18], modification of the optical properties of the medium [19] and many others, sufficiently higher densities (with effective pressures of few torrs or higher) are necessary. In such experiments, the degree of molecular alignment is extracted from time-resolved optical birefringence measurements (TROB) [20,21]. TROB relies on the transient birefringence induced in the gas as the latter periodically attains anisotropic angular distributions. This birefringence is sampled by a weak probe pulse that propagates through the anisotropic medium and analysed for changes in its polarization. Typically reported birefringence values are $10^{-6} < \Delta n < 10^{-5}$ and are induced by a short laser pulse with intensity $10^{13} \sim 10^{14} \frac{W}{cm^2}$ [22–25]. The birefringence is directly related to the desired degree of molecular alignment, $\Delta n = \frac{3N\Delta\alpha}{4\varepsilon_0} \langle\langle \Delta \cos^2 \theta \rangle\rangle$ (1), where $N$ is the gas density, $\Delta\alpha$ is the polarizability anisotropy ($\Delta\alpha = \alpha_\parallel - \alpha_\perp$ the difference in polarizability components parallel and perpendicular to the molecular axis) and $\langle\langle \Delta \cos^2 \theta \rangle\rangle$ is the change in alignment averaged over the thermal ensemble. The birefringence is experimentally deduced using the relation: $\frac{\Delta I}{I} = \sin\left(\frac{\omega L}{c} \cdot \Delta n\right)$ (2), where $I$ and $\Delta I$ are measured experimentally ($\frac{\Delta I}{I}$ is the intensity modulation of the probe), $\omega$ is the optical frequency of the probe and $L$ is the length of interaction between pump and probe beams. In what follows we refer to the method for extracting the alignment factor as 'the conventional method'.

*Inherent problems of the method:*

From all of the above, in order to extract the alignment factor from the experimental data, one must accurately know the experimental parameters: $N$, $L$ and $\Delta\alpha$. While $\Delta\alpha$ can be found in spectroscopy databases (such as https://cccbdb.nist.gov/), may be calculated [26] or measured by other means [27,28], the variability in the reported values often exceed to 10-20% [29]. The experimental parameters $N$, $L$ often remain highly elusive depending on the exact experimental configuration . For example, in experiments on cold ensembles expanded from molecular jets [30], the density of the gas strongly depends on the distance from the nozzle at which the interaction takes place in addition to its geometry, the backing pressure etc. These parameters also govern the effective length of interaction and the density gradient across the molecular beam and altogether may result in large deviations in the extracted degree of alignment. In case of a static gas cell (with accurately known gas density and temperature), the length of interaction($L$), may still be inaccurately defined in both the crossed-beam or collinear pump-probe geometries. A typical length of interaction may be estimated as the Rayleigh range of the two collinear beams, but clearly this is merely an approximation. In the crossed beam geometry, the length of interaction is also only roughly estimated, and the intensity gradient of both beams as they cross imparts additional uncertainties. In addition, the time-resolved optical birefringence signal is a convolution of the alignment response of the gas and the probe pulse. Therefore, in order to accurately obtain the birefringence signal one should deconvolve the signal with respect to well-characterized probe pulse duration. We note once more that in many experiments that aim to explore the rotational dynamics, the exact degree of alignment obtained may not be of much importance and indeed in many cases, it is not reported at all. Whether since it is hardly accessible (as described above), or just not important enough - one can only speculate.

In this letter we propose and demonstrate the use of the rotational echo response as an intrinsic observable for the experimental calibration of the alignment factor. The method is decoupled from both the interaction length and the gas density and relies on the unique dynamical features of rotational echoes [31–33] and specifically the oscillatory dependence of the echo amplitude with the intensity of the rephasing (2$^{nd}$) pulse.

Alignment echoes are rotational responses of molecular rotors, induced by two, time delayed laser pulses and have attracted much attention recently as they enable characterization of the decoherence rate in gas phase molecular ensembles and demonstrate rich and interesting coherent dynamics [31–40]. Very briefly, the first pulse (applied at t=0) creates a coherent rotational wavepacket that evolves under field-free conditions. The second pulse (applied at $t = \Delta\tau$) effectively reverses the wavepacket dynamics which results in the rephasing of the rotational wavepacket at $t = 2\Delta\tau$ that manifest in anisotropic angular distribution that was induced by the first pulse, only in the reversed time-evolution. We have recently studied the dependence of alignment echoes on $\Delta\tau$ and on the pulses' intensities [31,32] and have found that unlike echo in a two-level system (as in spin-echoes [41]), multi-level molecular rotors manifest much richer and more complicated echo responses that depend on both $\Delta\tau$ and the intensity of the rephasing pulse interweaved together. Specifically we have shown that for a fixed delay $\Delta\tau$ between the pulses, the echo signal amplitude ($S_{echo}$) depends on the intensity of the second pulse ($P_2$) as $I_{echo} \propto sin^2(a \cdot P_2)$. While this oscillatory dependence imposes hurdles to the implementation of echo spectroscopy in molecular rotors, the ability to identify and locate the maximal echo signal is a key feature for the proposed alignment factor calibration scheme as described hereafter.

Figure 1a shows a typical time-resolved optical birefringence measurement of carbonyl-sulfide (OCS) gas sample induced by two short laser pulses (~100fs pulse duration). The alignment signals induced by each pulse selectively are marked in the figure as well as the echo signal induced by the action of both pulses (here $\Delta t \cong 16 ps$ and the echo signal observed at $2\Delta t \cong 32 ps$). The Echo amplitude, $S_{echo}$ is quantified as the peak-to-peak amplitude as marked in the figure.

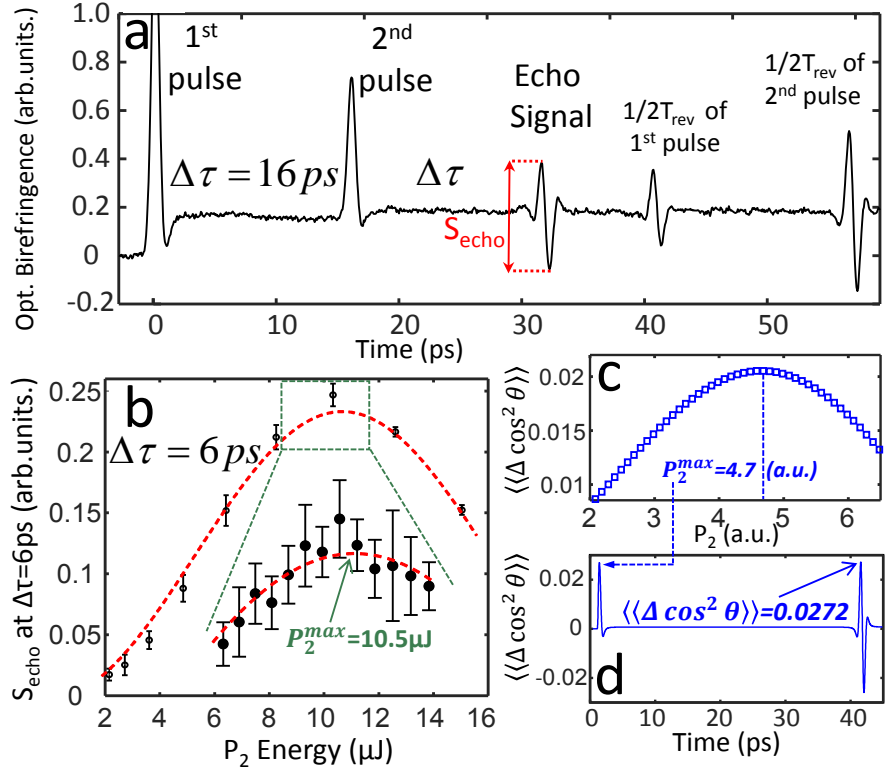

**Figure 1:** Methodology. (a) time-resolved scan of the rotational dynamics induced in 30torr OCS gas by two laser pulses with $\Delta \tau = 16 ps$ apart. The resulting signals are marked in the figure. (b) Measurement of the echo amplitude ($S_{echo}$) as a function of the 2$^{nd}$ pulse energy from which the experimental $P_2^{max}$ (=10.5 ± 0.5µJ ) is obtained. The delay between the pulses was set to $\Delta \tau = 6 ps$. (c) Simulated results of figure (b) in arbitrary units of the simulation for the case of figure (b) to obtain the simulated $P_2^{max}$. (d) Simulated alignment induced by the $P_2^{max}$ found in figure(c) as a single pulse resulting in $\langle\langle \Delta \cos^2 \theta \rangle\rangle = 0.0272$. From this example we deduce that a single pulse with $10.5 \pm 0.5 \mu J$ induces an experimental $\langle\langle \Delta \cos^2 \theta \rangle\rangle = 0.0272$ at the 1/2 $T_{rev}$.

Figure 1b depicts the experimental echo signal amplitude (peak-to-peak) from OCS, obtained with two pulses with $\Delta \tau = 6 ps$ as a function of the 2$^{nd}$ pulse energy, showing the sinusoidal squared dependence described above and in [32]. A finer scan of $P_2$ pulse energy (expanded by the green dashed lines in Figure 1b) provides accurate identification of $P_2^{max}$ i.e. the energy of the 2$^{nd}$ pulse that yields a maximal echo signal and will serve as an anchoring point in our calibration. Next, we performed a set of simulations for the echo response of OCS with $\Delta \tau$ (= $6ps$) and varying $P_2$ energies (intensities). The simulated $S_{echo}$ vs. $P_2$ energy is depicted in Figure 1c from which we extract the simulated $P_2^{max}$=4.7 (in the arbitrary units of the simulation) as marked by the blue dashed line. Using this value we calculate the degree of alignment that is induced by the simulated $P_2^{max}$ pulse when applied alone, resulting in $\langle\langle\Delta\cos^2\theta\rangle\rangle = 0.0272$ at the half revival as shown in Figure 1d. With the above information at hand, we return to the experimental setup, block the $1^{st}$ pulse and induce molecular alignment with the experimental $P_2^{max}$ (10.5 µJ) alone. The experimental alignment peak at $1/2T_{rev}$ is now calibrated to $\langle\langle\Delta\cos^2\theta\rangle\rangle = 0.0272$. In fact, since the value of $\langle\langle\Delta\cos^2\theta\rangle\rangle = 0$ is readily provided by the measurement (with both pump beams blocked), one is able to obtain a decent calibration even with only two data points obtained. The proposed method relies on the ability to experimentally identify the $P_2^{max}$ value in a two pulse echo experiment. Thus, it is crucial that $P_2^{max}$ is decoupled from $P_1$ intensity. This has been verified both experimentally and theoretically and reported in our previous work (see Figure 3b in [32] and associated text).

Note that in Fig.1b and associated text, we refer to the pulse energy (in [µJ]) instead of characterizing our pulse by its intensity (in [W/cm$^2$]). We deliberately report the pulse energy since our calibration technique does not require knowledge of the pulse intensity (for both the experiment and the simulation) and since potential users of the technique are also expected to characterize their excitation pulse by measuring the power of the excitation beam just like we do. Furthermore, the measured pulse energy can directly translate to pulse intensity since the beam diameter, its focusing parameters and pulse duration remain unchanged throughout our experiment.

To further improve the proposed calibration method, one would naturally need to obtain several calibration points that span a range of alignment factors induced by the $2^{nd}$ pulse. This can be done by repeating the experimental scheme described above for different delays between the two pulses.

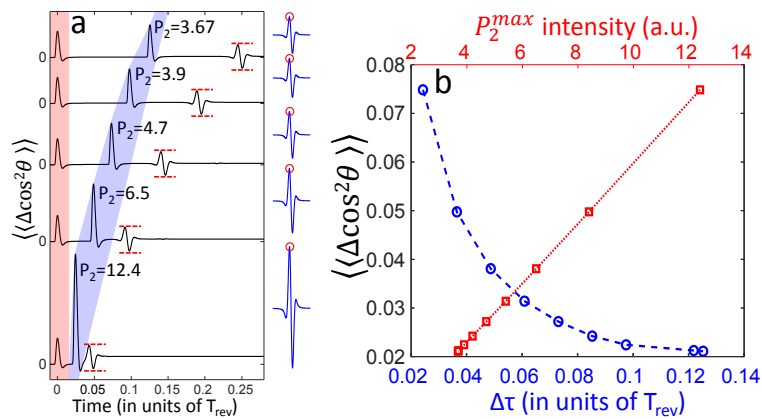

**Figure 2:** Construction of the calibration curve. (a) Simulated alignment responses for five different delays between pulses (in units of $T_{rev}$). The first pulse intensity is kept fixed (highlighted in red) and the second pulse intensity was numerically scanned to find $P_2^{max}$ for each of the delays separately (highlighted in blue, given in the arbitrary units of the simulation). On the right to each panel we plot the simulated alignment at $1/2T_{rev}$ induced by each of the $P_2^{max}$ selectively. (b) Calibration curve derived from the simulated results of (a). The alignment at the peak of the $1/2T_{rev}$ signals vs. the delay (blue curve) and $P_2^{max}$ (red curve). Note that we have simulated few more delay points to construct figure (b) and are not shown in fig. (a).

Figure 2a depicts simulated alignment responses for five different delays between the two pulses. The intensity of the first pulse was kept fixed in all of the simulated scans. For each delay $\Delta\tau$ between pulses we performed a series of simulations with varying P$_2$ intensities (as in Fig.1c, not shown here) from which we extract $P_2^{max}$ i.e. the P$_2$ intensity that gives rise to the maximal echo response (shown in Fig.2a). Note that the maximal echo signals are of equal amplitudes regardless of the delay [32,33]. The extracted $P_2^{max}$ values were fed back into the simulation as a single pulse excitation, the alignment signals at the 1/2T$_{rev}$ (blue signals in Fig.2a) obtained and their peak values (red circles) recorded. The latter are plotted in Fig.2b against $\Delta\tau$ between pulses (blue circles, x-axis at the bottom) and against $P_2^{max}$ (red squares, x-axis at the top). Thus, Fig.2b serves as a calibration curve, providing the association between the experimental $P_2^{max}$ values (obtained in echo measurements) and simulated $\langle\langle\Delta\cos^2\theta\rangle\rangle$ values. We note that since the dependence of $P_2^{max}$ on $\Delta\tau$ is an oscillatory function with a period of $\frac{1}{4}T_{rev}$ [32], the relevant range for the calibration curve is therefore $0 < \Delta\tau < \frac{1}{8}T_{rev}$.

*General aspects of the calibration method:*

The simulations shown in Fig.2 were performed for OCS gas (with B=0.203cm$^{-1}$) at ambient temperature (300K) and the pulse envelope taken as Gaussian with $I_1(t) = P_1 \times \exp[-t^2/\sigma^2]$, $I_2(t) = P_2 \times \exp[-(t-\Delta\tau)^2/\sigma^2]$, both with a full width half maximum (FWHM) of 100fs (as in our experiment). Linear molecules like OCS are modelled as quantum mechanical rigid rotors and their dynamics calculated by numerically propagating the density matrix $\rho$ in time via the Liouville-Von Neumann equation $\frac{\partial\rho}{\partial t} = -\frac{i}{\hbar}[H,\rho]$ where $H = \frac{\hat{L}^2}{2I} + \hat{V}$. The interaction term for the nonresonant rotational excitation is given by $\hat{V} = -\frac{1}{4}\Delta\alpha|E(t)|^2\cos^2\theta$.

We note that the results in Fig.2b are independent of P$_1$-the intensity of the first pulse and depend only on P$_2$-the intensity of the second pulse and the delay between the two pulses, $\Delta\tau$, given in units of the quantum revival period for generality. The simulated P$_2$ intensities (given by the red x-axis on the top of Fig.2b) are provided in the arbitrary units of the simulation, i.e. do not require prior knowledge of the molecular polarizability anisotropy ($\Delta\alpha$) or the experimental pulse intensity $|E_{(t)}|^2$ (the latter strongly depends on the overlap of the pump and probe beams at the interaction region, on the exact focusing parameters and the position of the focus). Instead, in order to use Fig.2b for calibration of the experimental $\langle\langle\Delta\cos^2\theta\rangle\rangle$, one needs to identify few experimental $P_2^{max}$ values from echo experiments performed with different $\Delta\tau$'s and associate them with $\langle\langle\Delta\cos^2\theta\rangle\rangle$ (as shown by the red curve) to replace the simulated P$_2$ intensities with the experimental ones (e.g. by measuring P$_2$ with a power meter). Once completed, a linear calibration curve of $\langle\langle\Delta\cos^2\theta\rangle\rangle$ vs. P$_2$ (in units of [Watt] or [Joule]) is obtained. Note that the calibration map does not depend on the elusive gas density (N) and the interaction length (L) since it relies solely on the intrinsic rotational echo response of the molecules. Furthermore, the experimental alignment signal (measured by the conventional method) is convolved with the probing pulse intensity profile. Thus, in order to extract $\langle\langle\Delta\cos^2\theta\rangle\rangle_{(t)}$ one must deconvolve the time-resolved signal with respect to the probe pulse. The proposed scheme inherently overcomes the need for deconvolving the alignment from the raw signal since it provides direct 'tagging' of the detected signal with its corresponding value of $\langle\langle\Delta\cos^2\theta\rangle\rangle$.

### *Comparison with the conventional characterization method*

To test the calibration scheme described above we performed an experimental comparison with the conventional method. Since the conventional method requires knowledge of the gas density (which is readily available in our static gas pressure cell) and the length of interaction (that is hardly accessible in collinear beams geometry), we modified our setup to cross-beams geometry (where the collinear pump beams are crossed by the probe beam at a small angle) in which we can experimentally determine the length of interaction as shown in Fig. 3. To compensate for the low signal level (owing to the short length of interaction), we used a sample of 90 torr carbon disulfide gas ($CS_2$) chosen for its large polarizability anisotropy $\Delta\alpha = 8.74 \text{Å}^3$ (taken from [42], however note that different sources report values ranging from $8 - 9.3 \text{ Å}^3$). The interaction length at the beams crossing was measured by polarization Kerr gating with a $180 \mu m$ thick glass slab as the Kerr medium, mounted on a linear stage and translated along the interaction of the two beams. The change in refractive index of the kerr medium is given by $n = n_0 + n_2 I_{pump}$ with $I_{pump}$ the intensity of the pump pulse, namely, $\Delta n \propto I_{pump}$. The birefringence induced in the Kerr medium is sampled via the same time-resolved optical birefringence setup described above and the detected signal is given by $\frac{\Delta I}{I} = \sin\left(\frac{\omega L}{c} \cdot \Delta n\right)$. With our pump pulse intensity significantly reduced to avoid damaging the glass Kerr medium and the short length of interaction dictated by the width of the glass (L = 180μm), the maximal modulation obtained was $\frac{\Delta I}{I} \sim 0.16$, and directly characterize the experimental length of interaction.

Figure 3a depicts the depolarization signal as a function of delay between pulses and position of the glass, the maximal signal in each position of the Kerr medium (depicted by the dashed red line) is fitted to a Lorenzian (Fig. 3b) with a full width half maximum of $789 \pm 87 \mu m$.

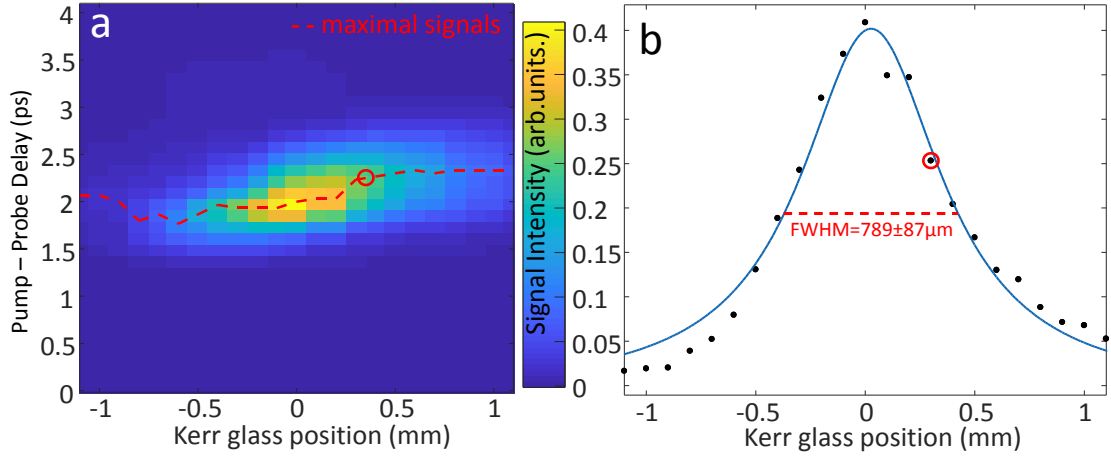

**Figure 3:** Kerr gate measurement results with a 180μm glass slide as the Kerr medium. (a) Probe depolarization measurement at different pump-probe delay and positions of the glass slide (the latter is mounted on a stage and scanned along the beam propagation direction). (b) a cross section along the maximal signal points in Fig.3a, depicted by the dashed red line. The length of interaction is extracted from the Lorenzian fit with FWHM of $789 \pm 87$μm.

Next, we performed the suggested experimental scheme for three different delays between pulses $\Delta\tau = 4\text{ps}, 10\text{ps}, 16\text{ps}$. For each delay we extracted the $P_2^{max}$ value (4.4 μJ, 2.3 μJ, 2.0 μJ respectively), applied each pulse alone to align the $CS_2$ ensemble and recorded the modulation $\frac{\Delta I}{I}$ values (0.127±0.017, 0.044±0.014, 0.027±0.015 respectively) at the $1/2T_{rev}$ alignment peak. Using $\langle\langle\Delta\cos^2\theta\rangle\rangle = \frac{\Delta I}{I} \cdot \frac{c}{\omega} \cdot \frac{4\varepsilon_0}{3N\Delta\alpha L}$ (3) (where we use the small angle approximation $\frac{\Delta I}{I} \approx \frac{\omega L}{c} \cdot \Delta n$) and the measured interaction length $L = 789 \pm 87$μm (Fig. 3) we extract the values of $\langle\langle\Delta\cos^2\theta\rangle\rangle_{conventional} = 0.043 \pm 0.008, 0.015 \pm 0.005, 0.009 \pm 0.005$ respectively. The simulated alignment factors using the proposed calibration method yielded $\langle\langle\Delta\cos^2\theta\rangle\rangle_{proposed} = 0.0491, 0.0211, 0.0161$ for the three delays respectively. We attribute the discrepancy between the two methods to the collisional decoherence experienced by the molecules and affects the conventional method, and was not taken into account in the proposed method- hence the lower values of the former compared to the latter. Thus, to take the collisional decoherence in account we performed a long scan of the 90torr, $CS_2$ sample and quantified the signal decay rate of $\gamma = 3.9 \times 10^{-3} ps^{-1}$ (using the same quantification metric as in [16,43]). This results in a decrease of 26% in the alignment signal at the $1/2T_{rev}$ $(\exp[-0.0039 ps^{-1} \times 76 ps] \cong 0.74)$ from its simulated value. Thus, the corrected $\langle\langle\Delta\cos^2\theta\rangle\rangle_{proposed}$ becomes (0.0365, 0.0157, 0.0120), in good agreement with the experimental values $(0.043 \pm 0.008, 0.015 \pm 0.005, 0.009 \pm 0.005)$ and as clearly shown, within the uncertainty of the conventional method.

*Conclusions*

We have utilized the non-monotonic, oscillatory behavior of rotational echoes to obtain intrinsic, self-contained anchor points for calibrating the laser-induced alignment factor. The proposed scheme is decoupled from the molecular polarizability, the interaction length and the gas density, that are hardly accessible in most of the configurations used in rotational dynamics experiments. We note however, that the calibration of the alignment factor does depend on temperature (the initial population distribution) through the simulated dynamics that can be extracted from the time-resolved alignment scan with a single excitation pulse [44] even if the density at the interaction region remains unknown. The technique can be applied in virtually all experimental setups as it depends on the intrinsic molecular echo responses and decoupled from the more elusive experimental parameters that are necessary for determining the degree of alignment via the conventional all-optical method. The proposed scheme was tested by comparing to the conventional method with good agreement between the two (with cross-beam geometry, well-defined gas density etc.). However, we note that in more desirable experimental conditions (e.g. molecular jet-expansion, collinear beam geometry to accommodate measurements of dilute gas samples) the direct comparison between methods is effectively impossible due to the uncertainties in the experimental parameters necessary for carrying out the conventional alignment characterization method. Such situations further reflect the dramatic advantages of our intrinsic, self-contained method for alignment factor characterization.

**Acknowledgements**

The authors acknowledge support of the Wolfson Foundation (Grants No. PR/ec/20419 and PR/eh/21797), the Israel Science Foundation – ISF (Grants No. 1065/14, 926/18, 2797/11 and by INREP—Israel National Research Center for Electrochemical Propulsion.